\documentstyle[prl,aps]{revtex} 
 
\newcommand{\bq}{\begin{equation}} 
\newcommand{\eq}{\end{equation}}

\begin{document} 
 
\draft 
\twocolumn[\hsize\textwidth\columnwidth\hsize\csname @twocolumnfalse\endcsname 
 
\title{ Magnetic clusters formation in Li$_{1-x}$Ni$_{1+x}$O$_2$ compounds :  
experiments and numerical simulations. } 
\author{ D. Mertz, Y. Ksari, F. Celestini*, J. M. Debierre, A. Stepanov} 
\address{ 
Laboratoire MAT\'eriaux Organisation Propri\'et\'es CNRS,  Universit\'e d'Aix-Marseille III \\ 
Facult\'e des Sciences de Saint-J\'er\^ome, Case 151, 13397 Marseille Cedex 20, France. } 
\author{ and C. Delmas} 
\address{ 
Institut de Chimie de la Mati\`ere Condens\'ee de Bordeaux CNRS, \\ 
Ecole Nationnale Sup\'erieure de Chimie et Physique 
de Bordeaux,\\87 Avenue du Docteur Albert Schweitzer, 33608 Pessac Cedex, France \\ } 
\date{Draft - Mars/30/1999 }

\maketitle

\begin{abstract} 
The magnetic properties of Li$_{1-x}$Ni$_{1+x}$O$_2$ compounds with $x$ ranging between $0.02$ 
and $0.2$ 
are investigated. Magnetization and ac susceptibility measured at temperatures between 
$2 $ K and  $300$~K reveal a high sensitivity to $x$, the excess Nickel concentration. 
 We introduce a percolation model describing the formation of Ni clusters and use
  an Ising model to 
simulate their magnetic properties. Numerical results, obtained by a Monte-Carlo technique,
 are compared 
to the experimental data. We show the existence of a critical concentration, 
$x_c = 0.136$, locating 
the Ni percolation threshold. The system is superparamagnetic for $x<x_c$,  
while it is ferrimagnetic for $x>x_c$. The $180^\circ$ \hbox{ Ni--O--Ni}  inter-plane 
super-exchange coupling $J_\perp \simeq -110K$ is confirmed to be the predominant magnetic
 interaction.   
 From the low temperature behavior, we find a clear indication of a $90^\circ$ \hbox{Ni--O--Ni} 
 intra-plane   
antiferromagnetic interaction $J_\parallel \simeq -1.5K$ which implies magnetic frustration. 
\end{abstract} 
 
\vskip2pc] 
 
\narrowtext 
\section { Introduction} 
Conduction and magnetic properties of Li$_{1-x}$Ni$_{1+x}$O$_2$ compounds have long 
been  studied  in relation with their crystallographic structure 
\cite{Dyer,Goode}. Two essential  
conclusions emerge from these studies. First, due to its lamellar structure allowing  
Lithium desintercalation/intercalation, 
LiNiO$_2$ is a good candidate as an electrode material in rechargeable 
Lithium-ion batteries \cite{these,Delmas}.  
Unfortunately, excess Nickel in the Lithium planes usually alters considerably 
the intercalation properties. Several synthesis ways have thus been developed in order to 
approach the stoichiometric composition \cite{Dutta}. Second, even if magnetic properties 
are known 
to be very sensitive to the structure, the actual relationship is not fully 
understood and still under debate. In this paper, we shall focus our attention on
 this latter point.\\

For large values of $x$, Li$_{1-x}$Ni$_{1+x}$O$_2$ keeps the  
rock-salt structure of NiO, in which Oxygen ions occupy one face-centered-cubic
 (fcc) lattice, 
while both  
Nickel and Lithium are randomly distributed over a second fcc frame 
\cite{Dyer,Goode,Goodebis,Bronger}. 
 As $x$ is decreased, the material leaves its cubic structure to adopt an
  hexagonal one in which 
the metal ions segregate into alternate Lithium and Nickel (111) planes, 
respectively denoted  
$(111)_{Li}$ and $(111)_{Ni}$. A lattice-gas model \cite{Reimers} has demonstrated 
that the transition occurs  
at a critical concentration $x^* \simeq 0.38$, in good agreement with X-ray 
diffraction experiments. 
When $x$ tends to $0$, the segregation tends to be complete and the almost  
stoichiometric compound shows an alternation of triangular Ni and Li planes 
(Fig.\ref{oxygen}). 
In this structure, a Ni plane is "magnetically isolated" by two adjacent Li 
planes so that LiNiO$_2$ can be 
considered as a model material for a perfect triangular two-dimensional (2D) magnet. 
Depending on the ferromagnetic 
or antiferromagnetic (AF) nature of the $90^\circ$ Ni--O--Ni magnetic coupling, 
the system could then be respectively ferromagnetic or frustrated at low temperatures. 
 Hirakawa {\em et al.} \cite{Hira} found evidence of a quantum liquid state, 
 as it 
is suggested by  Anderson for a triangular lattice  
Heisenberg antiferromagnet \cite{Anderson}. 
 Later, Kemp {\em et al.} \cite{Kemp} interpreted their experimental results 
as a consequence of a ferromagnetic intra-plane  interaction. Until now the 
question of the  
intra-plane coupling thus remains a controversial subject \cite{Yamaura,Bajpai}.  
For $x>0$, two other magnetic interactions should be considered. Charge 
compensation on Nickel 
implies that a Ni$^{2+}$ ion sitting in a $(111)_{Li}$ plane has five Ni$^{3+}$ 
and one Ni$^{2+}$ ions as nearest neighbors (NN) in the adjacent 
$(111)_{Ni}$ planes.  
The corresponding magnetic interactions are denoted $J_{\perp}$ 
and $J_{\perp}'$ respectively (Fig. \ref{Struct}). 
If $J_{\perp}'$  is known to be strongly antiferromagnetic, no direct 
conclusions can be made  
on the sign of $J_{\perp}$ \cite{Goode}.The existence  of these 
inter-plane interactions,  
due to the Anderson super-exchange \cite{And.super} and acting in the direction 
 perpendicular to the (111) planes, tends to destroy the 2D magnetic behavior
  expected in  
 the stoichiometric compound. This is 
illustrated in Fig. \ref{Struct}(a), where we show a magnetic Ni cluster 
containing one ion located in a 
$(111)_{Li}$ plane and six nearest-neighbors (NNs) located in the two 
adjacent $(111)_{Ni}$ planes. 
The two fundamental spin states, depending on the sign of $J_{\perp}$, 
are sketched in Fig. \ref{Struct}(b). 
Let us note finally that Kuiper {\em et al.} \cite{Kuiper,van-Elp} recently 
discussed the possibility that substitution of Li by Ni can give rise to a 
charge compensation on  
Oxygen rather than on Nickel. The resulting Ni$^{3+}-$O$^{-}$ exchange 
interaction being two orders  
of magnitude larger than the  Ni$^{3+}-$O$^{2-}-$Ni$^{2+}$  
super-exchange, this point should be easily tested by magnetic measurements.  
  
The aim of this study is to investigate the magnetic behavior of the
 Li$_{1-x}$Ni$_{1+x}$O$_2$ family 
through a comparison between 
experimental and numerical simulation results. We consider here the compositions 
$x=0.02, 0.07, 0.08, 0.14$ and  
$0.2$, lying in the 
region where metal ions segregate in the $(111)_{Li}$ and $(111)_{Ni}$ planes.  
In section II, we describe the sample preparation and characterisation procedure. 
We also give experimental results 
for the magnetization and ac susceptibility as functions of temperature.  
In section III, a percolation approach is used to simulate the formation of the 
Ni spin clusters. The magnetic properties of the disordered system 
are then calculated by performing Monte Carlo simulations in an Ising-like model. 
We finally present our discussion and conclusions in section IV:  we 
are able to confirm 
that the inter-plane interactions are two order of magnitude greater than 
the intra-plane ones, and present arguments in favor of an AF intra-plane 
interaction  
leading to magnetic frustration. The scenario consisting in charge
 compensation on oxygen 
is ruled out for the low concentrations examined in this paper.
 We finally  demonstrate the existence of a critical 
concentration, $x_c=0.136$,  
for the extra Ni ions. For $x<x_c$, the system behaves as a superparamagnet, 
while for $x>x_c$ it is a ferrimagnet.

\section{Experiments} 
 
Fifty-mg samples of Li$_{1-x}$Ni$_{1+x}$O$_2$ powder, prepared as reported in  
Ref. \cite{Delmas-prep}, were placed in Pyrex tubes, 
$4$ mm in diameter, filled with Helium gas. The amount of extra Nickel
 was previously  
determined by Rietveld analysis in the $\bar{R3m}$ crystallographic space group. 
The magnetic measurements were carried out using an Oxford Instrument Maglab$2000$ 
cryostat equipped  
with a $7$~T superconducting magnet and a variable temperature insert 
allowing temperature scans between $1.8$ and $450$ K. During experiments 
at constant field, the magnet was set in the persistent mode. 
 
The measurement system comprises an excitation coil for ac susceptibility, with two inner 
compensated coils, used either for ac susceptibility or for dc moment measurements. 
The dc moment and ac susceptibility signals are measured respectively with a 
Keithley $2001$ voltmeter  and with a SR$830$ lock-in-amplifier. 
The geometry of the system allows a sensitivity of $10^{-5}$ (dc moment) and $10^{-8}$ emu  
(ac susceptibility). 
  
The static properties were first investigated  under external  
field of $2$ and $5$ T as a function of $x$.  
These field values are sufficiently large to suppress the ferromagnetic domains  
due to dipole-dipole interactions and to make negligible the effect of magnetic anisotropy 
which reaches $2.5$~T in this compound \cite{Barra}. In these 
conditions, the system is only subject 
to local exchange interactions in which we are interested. 
The data were corrected for demagnetizing effects using 
a model of spherical thin grains. The correction is only significant below $100$ K 
for the largest $x$ values and does not exceed $5\%$  at $3$ K. 
In Fig. \ref{mag_lab}, we represent  the magnetization curves 
obtained under $2$ T. It is clearly seen that 
the magnetic response of the Li$_{1-x}$Ni$_{1+x}$O$_2$ family is strongly influenced  
by the amount of extra Nickel ions. The important point is that the topology of the  
experimental curves changes with $x$. This suggests that, as $x$ changes,  
the system undergoes 
a transition between distinct macroscopic magnetic behaviors probably related  
to Ni clusters formation. Indeed, the lowest ($x<0.14$) and 
highest ($x>0.14$) concentrations present respectively superparamagnetic and
ferromagnetic like behavior.  
To illustrate and quantify this point we measured the ac susceptibility 
of the samples at a 
frequency of $10$ kHz, with an excitation field of $3$ Oe. In Fig. \ref{khi_lab}, we  
show the data obtained for $x=0.02$, $x=0.14$ and $x=0.20$. The susceptibility curves 
display maxima  which are characteristic for ferromagnets at temperatures of
 $116$K and $168$K
respectively for $x=0.14$  and $x=0.20$. In contrast, for $x=0.02$, the 
temperature variation  
of $\chi_{ac}$ is rather monotonic. These observations confirm the
 superparamagnetic behavior of 
compounds with low extra Nickel concentrations, in agreement with previous neutron 
experiments \cite{Hira}.       
The second point concerns the magnetization of the different samples at 
the lowest temperatures.  
Assuming that all the magnetic moments in Li$_{1-x}$Ni$_{1+x}$O$_2$ are 
parallel, the magnetization 
should saturate to: 
\begin{equation} 
 M_s = \left( 1-x \right) N \mu _B g_1 S_1 + 2xN\mu_Bg_2 S_2, 
\end{equation} 
$N , \mu_B, g$ and $S$ denoting respectively the Avogadro's constant,
Bohr's magneton, Land\'e's factor and spin value. 
 The first term represents the contribution of Ni$^{3+}$ ions in the 
$(111)_{Ni}$ planes while the  
second one is related to Ni$^{2+}$ ions in $(111)_{Li}$  
and $(111)_{Ni}$ planes. This saturation should be reached when the Zeeman energy exceeds  
the interplane antiferromagnetic coupling.       
This cannot be reasonably the case for the moderate magnetic fields involved in this work.  
In Fig. \ref{mag_satur}, we plot the magnetization measured at $4.3$ K,  as a function of $x$,  
together with $M_s(x)$ (Eq. 1). 
We can see that the maximum saturation is not reached. 
 This is in good agreement with previous results 
of Rougier {\em et al.} \cite{these} who reported the absence of saturation 
for $x=0.02$, even at $11$ T. In addition, our results show that the measured magnetization  
presents a maximum at $x \simeq 0.14$.         
 In the light of the experimental results, it is difficult to conclude about the exact 
nature of the magnetic order. In the next section, we propose a simple model to investigate  
this point further.

\section{Model} 
 
Our model incorporates the different magnetic interactions described in Section I. 
The main ones ($J_\perp$ and $J_\perp'<0$ ) are the $180^\circ$ Ni--O--Ni  
super-exchange
coupling one Ni ion sitting in a $(111)_{Li}$ plane to its six NNs 
sitting in the adjacent $(111)_{Ni}$ planes. It is reasonable to assume that these two coupling 
constants are of the same order of magnitude. In the following we suppose  
$J_\perp=J_\perp'< 0$.  
It is obvious that changing  the sign of $J_\perp$ does not modify the magnetization, the 
susceptibility and other related quantities, since there 
is a compensation between the magnetic moments of the two Ni$^{2+}$ ions.  
 We also consider the $90^\circ$ Ni--O--Ni intra-plane  interaction $J_\parallel$. 
We make no  {\em a priori} assumption about the sign of $J_\parallel$.  
Recent neutron diffraction studies \cite{Pouillerie} on this material family 
($x \leq 0.20$) have shown that there is no 
Lithium in the Nickel planes.  
We verified this point with the help of the lattice gas 
structural model introduced by Reimers {\em et al.} \cite{Reimers}. 
For a composition $x=0.2$, which is the highest considered here, the quantity 
of Li ions in the   
$(111)_{Ni}$ planes is not significant (less than $2\%$) and can be reasonably 
neglected.  
Furthermore, we suppose that the excess Ni ions are randomly distributed 
 in the $(111)_{Li}$ 
planes. 
Since we assumed that the excess Ni ions are antiferromagnetically coupled 
to their six NNs, 
three-dimensional magnetic Ni  
clusters are formed in the system, as shown in Fig. \ref{Struct} (for the time being, 
we neglect intra-plane  coupling). 
For $x<x_c$ all these clusters have a finite size. At the critical 
concentration $x_c$ the  
largest cluster percolates and for higher concentrations it grows in size. 
This percolation transition strongly affects the  
magnetic properties. One may expect  different types of behaviors in samples
 with $x$ below, at, 
or above $x_c$. 
Numerical simulation gave $x_c\simeq 0.136$ for the percolation  
threshold. The cluster mass distribution $n_s$ was 
calculated at $x_c$ : it follows 
a power law $n_s \propto s^{-\tau}$ with a critical exponent $\tau = 2.19 \pm 0.01$. 
This last 
result indicates a regular three-dimensional percolation transition \cite{JM}, 
as was predicted 
in a dedicated 
study of percolation in multi-layered structures \cite{Dayan}. 
We note that a small amount of Nickel impurities can give rise to large clusters 
because each impurity is 
coupled to six other Ni ions. This is illustrated in Fig. \ref{Clu2} where 
we represent the Ni ions distribution 
in a sample with $x=0.02$. For such a low value of $x$, most of the magnetic 
clusters contain seven ions,  
but larger clusters coupling three successive Ni layers are also observed.

In order to calculate the magnetic properties, we use an Ising model in which  
each Ni ion has a spin $\sigma=\pm 1$. 
The choice of an Ising-type Hamiltonian is justified by intrinsically high 
magnetic anisotropy of Ni clusters in LiNiO$_2$ \cite{Barra}, and {\em a posteriori} by 
the good agreement found between the 
numerical results and the experimental data.  In our model, we neglect the differences 
between the Ni ions. Indeed, the $(111)_{Ni}$ planes are mostly composed of  N$i^{3+}$ ions 
in a low spin state $S=1/2$ 
while extra Nickels are Ni$^{2+}$ ions with $S=1$ \cite{Azzoni}. Two types of spin can be 
easily taken into account in the model, but we checked that it is not a 
crucial point, probably because   
we consider small concentrations of extra Nickel ions. 
 
 The Hamiltonian then simply  reads: 
\begin{equation} 
  H = -g\mu_B B \sum_i \sigma_i - J_\perp \sum_{<i,j>} \sigma_i \sigma_j - J_\parallel  
  \sum_{<i,j>} \sigma_i \sigma_j 
\end{equation} 
The first term describes the interaction with the magnetic field B. 
In the second term, $\sigma_i$ sits on a Ni in a $(111)_{Li}$ plane, 
and the $\sigma_j$'s sit on the six NN Ni ions in the adjacent $(111)_{Ni}$ planes. 
In the last term, the sum runs 
over all NN pairs in the same plane. 
For a given spin configuration, the Boltzmann factor is: 
\begin{equation} 
    \exp \left[ {K \left( b\sum_i \sigma_i 
    + {\rm sign}(J_\perp) \sum_{<i,j>} \sigma_i \sigma_j + j_\parallel  
 \sum_{<i,j>} \sigma_i \sigma_j \right) } \right] 
\end{equation} 
 
where $b=g\mu_B B/\left|J_\perp\right|$, $K=\left|J_\perp\right|/kT$,  
and $j_\parallel=J_\parallel/\left|J_\perp\right|$ are respectively 
the dimensionless field, the inverse temperature, and the ratio 
of intra- to inter-plane coupling constants. \\ 
The classical "single spin flip" Monte Carlo algorithm is not well suited 
to simulate this system 
because of critical slowing down around the transition and at low
 temperatures \cite{MC}.  
To avoid very long simulation times, we adapted Wolff's algorithm \cite{Wolff} 
to our system. This method largely compensates the correlation length 
increase when 
the temperature is lowered because the system evolves by flipping clusters 
of ordered spins.  
One Monte Carlo step (MCS) consists in constructing a Wolff cluster and then 
 trying to flip it.   
In our case, the construction of a cluster is made as follows: we randomly 
choose a first spin $\sigma_a$. 
All the inter-plane bonds (if any) between $\sigma_a$ and the
closest spins $\{\sigma_b\}$ 
are examined in turn. 
Those with $\sigma_a=\sigma_b$  are discarded; for the other bonds, 
spin $\sigma_b$ is added to the cluster with an  
arbitrary probability $p$. Now set $\{\sigma_a\}$ is replaced 
by set $\{\sigma_b\}$ and the procedure 
is repeated until an empty set is found. 
Choosing $p=1-e^{-2K}$, as in the classical Wolff algorithm, 
the expression for the flipping  
probability satisfying the detailed balance condition simplifies to  
\begin{equation} 
\min \left( 1, \exp \left[-2 K\left(  b M_c +  j_\parallel
 \Gamma_c \right)\right] \right) 
\end{equation} 
with 
\begin{equation} 
     M_c=\sum_{cluster} \sigma_i 
\end{equation} 
and 
\begin{equation} 
    \Gamma _c = \sum_{<i,j>_{surf}} \sigma_i \sigma_j 
\end{equation} 
The last sum runs over all the pairs of spins at the cluster surface, 
$\sigma_i$ being part of the cluster and $\sigma_j$ at the exterior.  
The exponential argument is the energy cost to flip the cluster: the 
first term is due to the field, the second to 
the intra-plane  interactions at the cluster surface. 
 
We considered systems of about $3000$ spins (9 pure $(111)_{Ni}$ planes,  
each  
containing $18 \times 18$ Ni ions, and $x\%$ excess Ni in the $(111)_{Li}$ 
 planes). 
Periodic boundary conditions 
were applied in the three directions. To check for finite size effects, we 
simulated a bigger system of approximately 30000 spins without significant 
changes in the  
results. For a concentration $x$, a 
spatial configuration was built up by randomly placing Ni excess ions in 
the Li planes. 
At each temperature, ranging 
from $300$ K down to $20$ K, 
we first equilibrated the spin system through about half a million MCSs and  
collected 
thermodynamic quantities 
during another million MCSs.  The same operation was repeated for several 
samples with different spatial configurations.  
Quantities such as magnetization $M$ and susceptibility $\chi$ , were finally 
obtained by averaging over the different samples. 
 
We first present the magnetization curves for samples with $x=0.02, 
0.14$ and $0.2$, in 
an external field $B=5$~T. The numerical results are given in Fig. \ref{MagB5}, 
together with the experimental ones.  
The value of the intra-plane interaction $J_\parallel$ is fixed to zero, 
while the antiferromagnetic inter-plane 
coupling is equal to $-110 $ K. This value, adjusted to obtain the best 
agreement between our model 
and the experimental data, is reasonably close to the values proposed 
in the literature for the $180^\circ$ Ni--O--Ni interaction \cite{NiO}. 
This estimate of $J_\perp$ rules out the possible presence 
of oxygen holes at these concentrations, since, as discussed above, this 
would give 
a much larger inter-plane interaction. 
The model reasonably reproduces the experimental magnetization curves, 
especially in the high temperature region. We can interpret our results in 
the following way. 
For $x=0.02$ the Ni clusters 
do not percolate and are  thus "magnetically isolated": they are at the origin of 
the superparamagnetic behavior observed 
both numerically and experimentally. For an extra Nickel concentration above the 
critical value ($x_c\simeq 0.136$), a magnetic 
cluster spans the sample. When the temperature is decreased, this magnetic cluster 
undergoes a second order transition  
to a ferrimagnetic state. Since the external field is high, the transition is 
broadened on the $x=0.14$ and $x=0.2$  
curves. We also calculated magnetization curves for the same concentrations but
 without  
external magnetic field. In this case, the sample with $x=0.02$ has no
 magnetization, 
as expected for a  
superparamagnetic system under zero applied magnetic field. On the other hand, 
in the samples with $x>x_c$, the   
transition is well marked with the appearance of a spontaneous magnetization 
at low temperature. 
 
 From the $M(T)$ curves, it is possible to extract values for the temperature $T_o$
locating the onset of magnetic ordering.  
For $x>x_c$, we first locate the inflexion point $I$. We then draw the line 
tangent to the curve at point $I$, and define $T_o$  
as the intercept with the $T$ axis. For $x<x_c$, there is no longer an inflexion point, 
since the system is superparamagnetic. However, we can still estimate the onset
 of magnetic ordering 
inside the isolated clusters. To do so, we define $T_o$ as the temperature 
at which magnetization reaches  
$7$ percent of its saturation value. For $x>x_c$, this definition approximately 
gives the same $T_o$ as the inflexion point method. In Fig. \ref{Tcrit},
 we represent 
the onset of magnetic ordering temperatures, $T_o$ deduced from  
the experimental and numerical $M(T)$ curves under an applied field of $2$ T. 
The good agreement confirms that our simple model permits an accurate description  
of the onset of magnetic ordering, thus providing information for the estimate  
of $x$.

Let us now examine more closely the low temperature magnetization.  
The disagreement between the experimental and numerical results can be traced 
to the different approximations made in the model. At first, 
the choice of Ising spins is certainly too crude and it affects more 
and more the shape of the calculated magnetization curve as $T$ is decreased. Secondly,
 we did not 
yet introduce the intra-plane interaction, known to be weaker than 
the inter-plane one, and  to influence the  
magnetization at lower temperatures. Assuming a strong AF inter-plane 
coupling $J_\perp$ and a weak intra-plane  coupling 
$J_\parallel$, we still have to discuss the sign of the intra-plane  coupling. 
We consider two different magnetic subsystems: the first one is composed 
of spins belonging to the ferrimagnetic clusters and the second one contains all the 
remaining spins which lie in the $(111)_{Ni}$ planes. We denote $\beta(x)$ the 
proportion of spins belonging to the clusters. We calculated 
$\beta(x)$ numerically for $x$ ranging between $0$  
and $0.2$: the results are well fitted by the polynomial $\beta(x)= 7x - 
22x^2 + 40.7x^3$. 
Assuming a ferromagnetic coupling, $J_\parallel \geq 0$,  at low temperature all 
the Ni$^{3+}$ ions  
in the $(111)_{Ni}$ planes adopt the field direction while the 
extra Ni$^{2+}$ spins  
adopt the opposite one. As one can see on Fig. \ref{Struct}(b) the 
spin $S=1$ on the extra Ni$^{2+}$ is exactly compensated by one of his 
NNs having a spin $S=1$ too. 
One can then expect a maximum ferrimagnetic magnetization 
\bq 
\frac{M_1}{N \mu_B g_1 S_1}=\left(1-x\right).  
\eq 
On the opposite, an AF coupling, $J_\parallel < 0$, leads to frustration 
in the second subsystem. In the case where this frustration is large enough to cancel 
the corresponding 
magnetization, only the ferrimagnetic clusters contribute to the low
 temperature magnetization. 
One then obtains  
\bq 
\frac{M_2}{N \mu_B g_1 S_1}=\left[\left(1+x\right)\beta\left(x\right)-2x\right] . 
\eq 
In Fig. \ref{mag_satur} we plot the two curves corresponding to $M_1(x)$ and $M_2(x)$ 
together with the experimental results for $B\!=\!2$ T at $T\!=\!4.3$ K. 
Since temperature is too high and magnetic field too low 
to ensure perfect saturation of the different subsystems, 
neither $M_1$ nor $M_2$ fit the experimental data on the whole range. We can however 
qualitatively discuss the two limits. $M_1(x)$ decreases linearly with $x$, in 
contradiction 
with the increase of magnetization observed at low $x$. Conversely, the $M_2(x)$
curve approximately follows the experimental data points up to a systematic vertical shift. 
Since Eq. (8) assumes that the second subsystem has no magnetization, 
this is consistent with an AF intra-plane  coupling ($J_\parallel < 0$).  
 
We thus performed additional simulations to examine the effect of the 
intra-plane  interaction $J_\parallel$. Our results  
show that, for a ferromagnetic  coupling ($J_\parallel \geq 0$), the magnetization 
curve changes concavity around $20$ K to reach the expected saturation value $M_1$ as  
$T\rightarrow 0$ K (Fig. \ref{In-plane}). For an AF coupling, ($J_\parallel<0$), 
the curves at low temperatures  
behave much more like the experimental ones: saturation  
is not reached because the AF coupling induces a frustrated state in the triangular 
$(111)_{Ni}$ planes. The best qualitative agreement  between simulations and 
experiments is found for 
$J_\parallel\simeq -1.5 $ K. Of course, this rough estimate of 
$J_\parallel$ is obtained for  
a given  value of $J_\perp$, determined from the high temperature 
behavior of the model with 
$J_\parallel=0$. For a better estimate, one would have to take 
both $J_\perp$, $J_\perp'$  and $J_\parallel$  
as free parameters, simulate several magnetization curves and 
fit them to the experimental data.

\section{Summary and discussion} 
 
In summary, we introduced a simple model to investigate the 
magnetic properties of non stoichiometric Li$_{1-x}$Ni$_{1+x}$O$_2$ 
layered compounds. The formation of three-dimensional (3D) Ni clusters 
around the extra Nickel ions was simulated by simple percolation of the Ni 
ions in the direction perpendicular to the metal layers. 
An Ising Hamiltonian was then used to compute the magnetic properties 
of the disordered system made of both 3D Ni clusters (magnetic 
coupling $J_\perp$) and 2D intra-plane  Ni clusters (magnetic 
coupling $J_\parallel$). The transition from a superparamagnetic 
to a ferrimagnetic behavior, observed experimentally as $x$ increases, 
is interpreted in our model as the onset of percolation for the 3D 
Ni clusters, at a critical threshold $x_c$. Our numerical estimate 
$x_c\simeq 0.136$, agrees nicely with the experimental evidence of 
magnetic ordering at $x=0.14$. 
 
The best fit of our model to the experimental  
data is found for $J_\parallel=-1.5K$ and  $J_\perp=-110K$. The assumption that
 $|J_\perp| \simeq |J_\perp'|$ is validated by  
the good agreement between the model and the experiments. The order of  
magnitude found for $J_\perp$ allows us to confirm magnetic coupling through
 the Ni-O-Ni super-exchange interactions.
 The possibility  of charge compensation on Oxygen
proposed for $x$ close to $1$  \cite{Kuiper}
is thus inadequate for the low $x$ values considered in this paper.
Another important outcome of our 
work is the relation found between the ordering temperature $T_o$
and the concentration $x$ of excess Nickel, which offers 
a new way to determine $x$ precisely. In fact the accurate 
determination of $x$ was overlooked for a long time.  
If new techniques \cite{Delmas-prep} permit now to determine  
the proportion of extra Ni ions with a reasonable accuracy, one should 
certainly be more  careful about 
older results for stoichiometric LiNiO$_2$, which very likely concern 
samples with a non zero effective value of $x$. 
 
Moreover, Bajpai {\em et al.} \cite{Bajpai} have recently reported the coexistence  
of ordered and random phases in the same sample, the proportion of each phase 
depending on the heat treatment during elaboration. The coexistence of 
two distinct phases may very well explain the anomaly observed at 240K 
in the ac susceptibility of some samples, a point currently 
investigated \cite{Andre_Nico}. 
 
Finally, the antiferromagnetic nature of $J_\parallel$ implies 2D magnetic 
frustration because the Ni planes have a trigonal symmetry. The magnetic properties 
of the system at very low temperatures ($T<\vert J_\parallel \vert$) 
would thus probably be better described in a 2D model with a random distribution 
of quenched spins due to the intra-plane coupling. We plan to investigate this 
point in the near future. \\ \\ 
 
*e-mail: celest@matop.u-3mrs.fr

\begin{figure} 
 
\vspace{1cm} 
\caption{ 
Perspective view of the Li$_{1-x}$Ni$_{1+x}$O$_2$ structure.  
} 
\label{oxygen} 
 
\vspace{1cm} 
\caption{ 
(a) Positions of the Nickel and Lithium ions in the LiNiO$_2$ structure (Oxygen 
ions are not represented). 
The presence of an extra Ni ion in a $(111)_{Li}$ plane induces the formation of 
a seven-ion Ni cluster and alterates the 2D behavior expected for 
stoichiometric LiNiO$_2$. 
(b) Depending on the sign of $J_\perp$, the two possible relative spin orientations 
are sketched.  
} 
\label{Struct} 
 
\vspace{1cm} 
\caption{ 
Temperature variation of the magnetization for five samples of the
 Li$_{1-x}$Ni$_{1+x}$O$_2$  family. 
} 
\label{mag_lab} 
 
\vspace{1cm} 
\caption{ 
Plot of the in-phase component of $\chi_{ac}$ 
for the $x=0.02$, $x=0.14$ and $x=0.20$ samples ($f=10$ kHz, $h_{ac}=3$ Oe). 
} 
\label{khi_lab} 
 
\vspace{1cm} 
\caption{ 
Magnetic moment measured at $T=4.3$ K and $B=2$ T as a function of the extra
 Nickel concentration.  
Squares represent our experimental data and 
diamonds are taken from Ref. [3]. The full line corresponds to 
the maximum saturation $M_s$. The dashed and dotted lines correspond  to   
calculated magnetizations for the different cases of intra-plane exchange
 interactions as defined in section III: 
respectively $M_1$ $(J_\parallel \geq 0)$ and $M_2$ $(J_\parallel < 0)$. } 
\label{mag_satur}

\vspace{1cm} 
\caption{ 
A spatial configuration of magnetic Ni ions in 
 Li$_{0.98}$Ni$_{1.02}$O$_2$. Lithium and Oxygen atoms 
are omitted for clarity. 
Grey spheres correspond to Ni ions belonging to ferrimagnetic clusters,
 white ones to other 
Ni ions. In the lower left-hand corner, one can see a cluster  
containing  more than one extra Nickel. 
} 
\label{Clu2} 
 
\vspace{1cm} 
\caption{ 
Reduced magnetization of Li$_{1-x}$Ni$_{1+x}$O$_2$ for $x=0.02$, $0.14$ 
and $0.2$, measured under an external field $B\!=\!5$ T. Full lines 
and symbols represent respectively experimental and numerical results. 
} 
\label{MagB5} 
  
\vspace{1cm} 
\caption{ 
Magnetic ordering temperature T$_0$, deduced from magnetization 
curves, as 
a function of $x$. Open diamonds correspond to numerical results, black 
circles to experimental ones. 
} 
\label{Tcrit} 
 
\vspace{1cm} 
\caption{ 
Reduced magnetization of Li$_{0.86}$Ni$_{1.14}$O$_2$ calculated for an 
applied field $B\!\!=\!\!5$ T, an antiferromagnetic 
coupling $J_\perp\!\!=\!\!110 $ K, and intra-plane  coupling 
$J_\parallel=1$ K, $0$ K, $-1$ K, $-2$ K and $-3$ K. 
} 
\label{In-plane} 
\end{figure}


\begin{references} 
 
\bibitem{Dyer} 
 L. D Dyer, B. S. Jr. Borie and G. P. Smith, J. Am. Chem. Soc. {\bf20}, 1499 (1954) 
 
\bibitem{Goode} 
 J. B. Goodenough, D. J. Wicham and W. J. Croft, J. Phys. Chem. Sol.
  {\bf5}, 107 (1958)  
  
\bibitem{these} 
A. Rougier, C. Delmas and G. Chouteau, J. Phys. Chem. Sol. {\bf 57}, 1101 (1996)and  
A. Rougier, Phd Thesis, Universit\'e de Bordeaux (1995). 
 
\bibitem{Delmas} 
C. Delmas, J.P. P\'er\`es, A. Rougier, A. Demourges, F. Weill, A.
 Chadwick, M. Broussely, 
F. Perton, Ph. Biensan and P. Willmann, J. Pow. Sources {\bf 68}, 120 (1997) 
 
\bibitem{Dutta} 
 G. Dutta, A. Manthiram, J. B. Goodenough and J. C. Grenier, J. Sol. Sta.
  Chem. {\bf96}, 123 (1992) 
   
\bibitem{Goodebis} 
 J. B. Goodenough, D. J. Wicham and W. J. Croft, J. Appl. Phys. {\bf29}, 382 (1958)  
  
\bibitem{Bronger} 
V. W. Bronger, H. Bade and W. Klemm, Z. Anorg. Allg. Chem. {\bf333}, 188 (1964)  

\bibitem{Reimers} 
W. Li, J. N. Reimers and J. R. Dahn, Phys. Rev. B {\bf 46}, 3236 (1992) and Phys. Rev.
 B {\bf 49}, 826 (1994)  
 

\bibitem{Hira} 
K. Hirakawa, H. Kadowaki and K. Ubukoshi, J. Phys. Soc. Jap. {\bf 9}, 3526 (1985) 

 
\bibitem{Anderson} 
P. W. Anderson, Mater. Res. Bull. {\bf 8}, (1973) 
 
\bibitem{Kemp} 
J. P. Kemp, P. A. Cox and J. W. Hodby, J. Phys. Cond. Mat. {\bf 2}, 6699 (1990) 
 
\bibitem{Yamaura} 
K. Yamaura and M. Takano 
J. Sol. State Chem. {\bf 127}, 109 (1996) 
 
\bibitem{Bajpai} 
A. Bajpai and A. Banerjee, Phys. Rev. B {\bf 55}, 12439 (1997) 
 
\bibitem{And.super} 
P.W. Anderson, Phys. Rev.  {\bf 79}, 350 (1950) 
 
\bibitem{Kuiper} 
P. Kuiper, G. Kruizinga, J. Ghijsen, G. A. Sawatzky and H. Verweij, Phys.
 Rev. Lett. {\bf 62}, 221 (1989) 
 
\bibitem{van-Elp} 
J. van Elp, H. Eskes, P. Kuiper, and G. A. Sawatzky, Phys. Rev. B {\bf 45}, 1612 (1992) 
 
\bibitem{Delmas-prep} 
A. Rougier, P. Gravereau and C. Delmas, J. Electrochem. Soc. {\bf 14}, 1168 (1996) 
 
\bibitem{Barra} 
A.L. Barra, G. Chouteau, A. Stepanov and C. Delmas, J. Mag. Mag. Mat.
 {\bf 177-181}, 783 (1998) 
 
\bibitem{Pouillerie} 
C. Pouillerie, J.P. P\'er\`es, E. Suard and C. Delmas, Solid State Ionics (Submitted) 

 
\bibitem{JM} 
see, e.g., D. Stauffer and A. Aharony, {\em Introduction to percolation theory}, 
(Taylor and Francis, London, 1992). 
 
\bibitem{Dayan} 
 I. Dayan, J. F. Gouyet and S. Harlin, J. Phys. A {\bf 24}, L287 (1991)  

\bibitem{Azzoni} 
C. B. Azzoni, A. Paleari, V. Massarotti, M. Bini and D. Casponi, Phys. Rev.
 B {\bf53}, 703 (1996)  
 
\bibitem{MC} 
K. Binder {\em Monte Carlo Methods in Statistical Physics}, 2nd Ed.
 (Springer Verlag, Berlin, 1986) 
 
\bibitem{Wolff} 
U. Wolff, Phys. Rev. Lett. {\bf 62}, 361 (1989) 
 
\bibitem{NiO} 
R. H. Kodama, S. A. Makhlouf and A.E. Berkowitz, Phys. Rev. Lett. {\bf 79}, 
1393 (1997) and references therein. 
 
\bibitem{Andre_Nico} 
A. Ghorayeb and N. Menguy, private communication. 
 
 
\end{references}
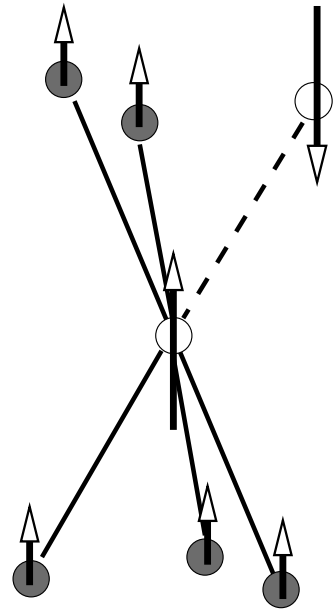
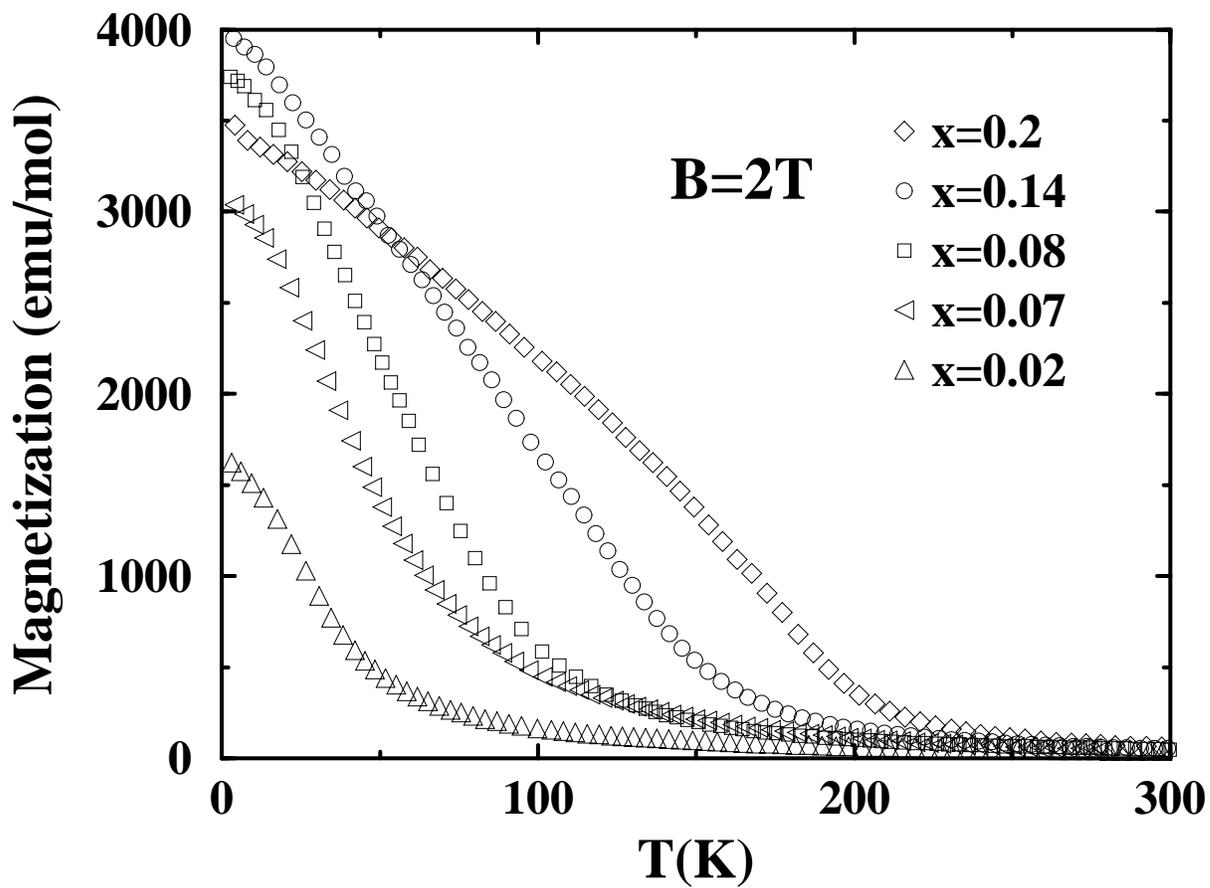
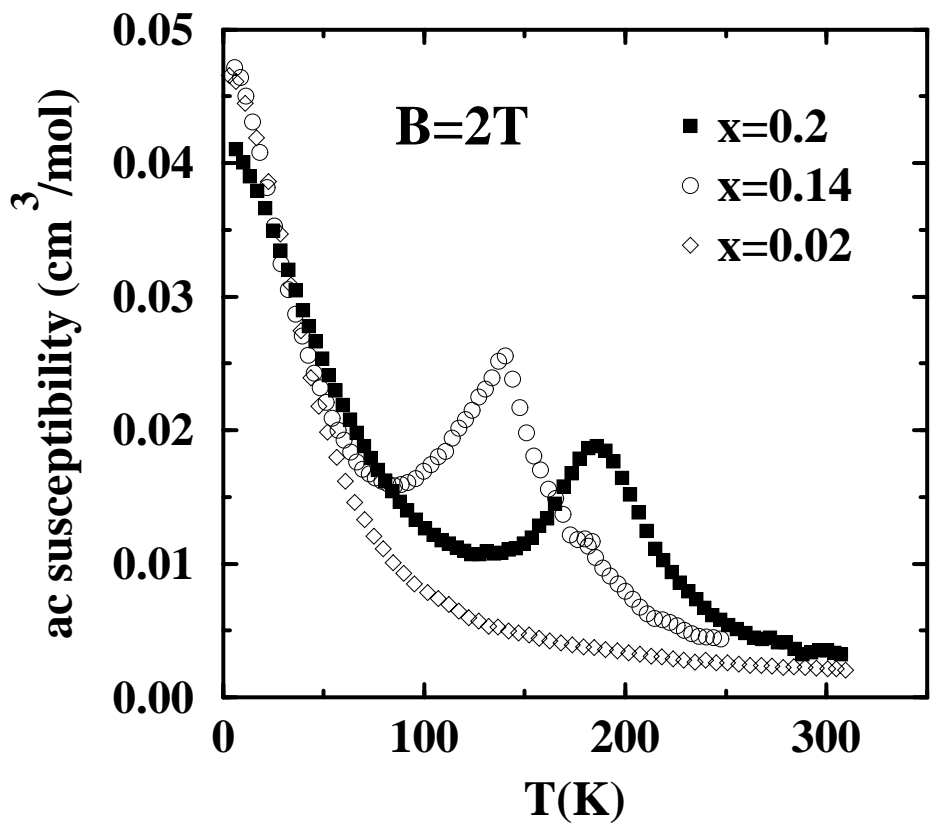
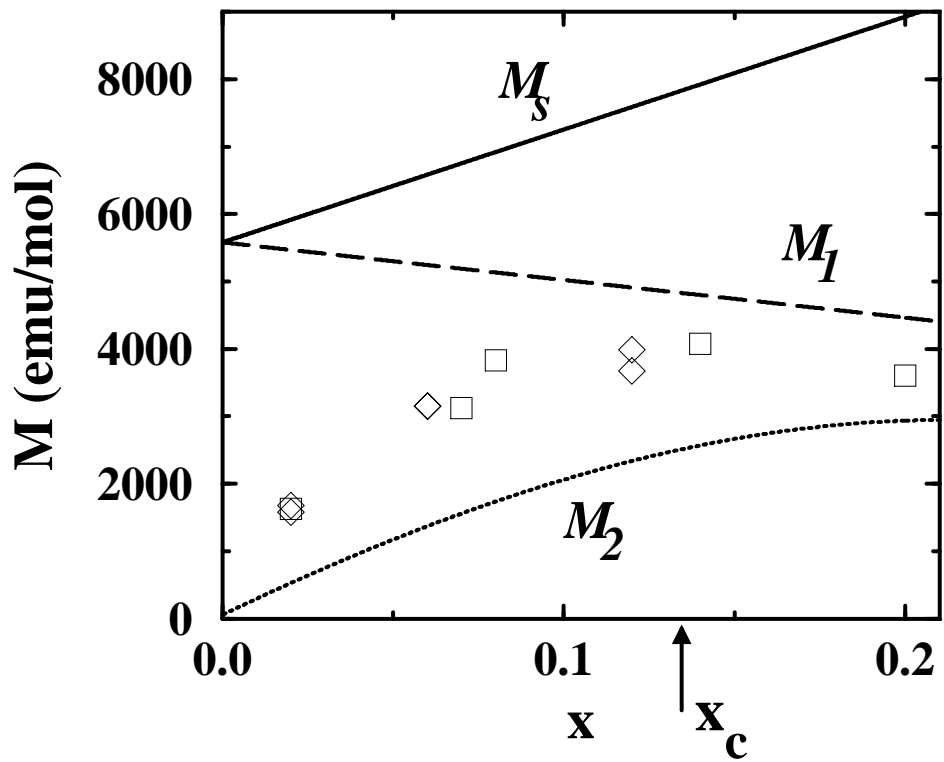
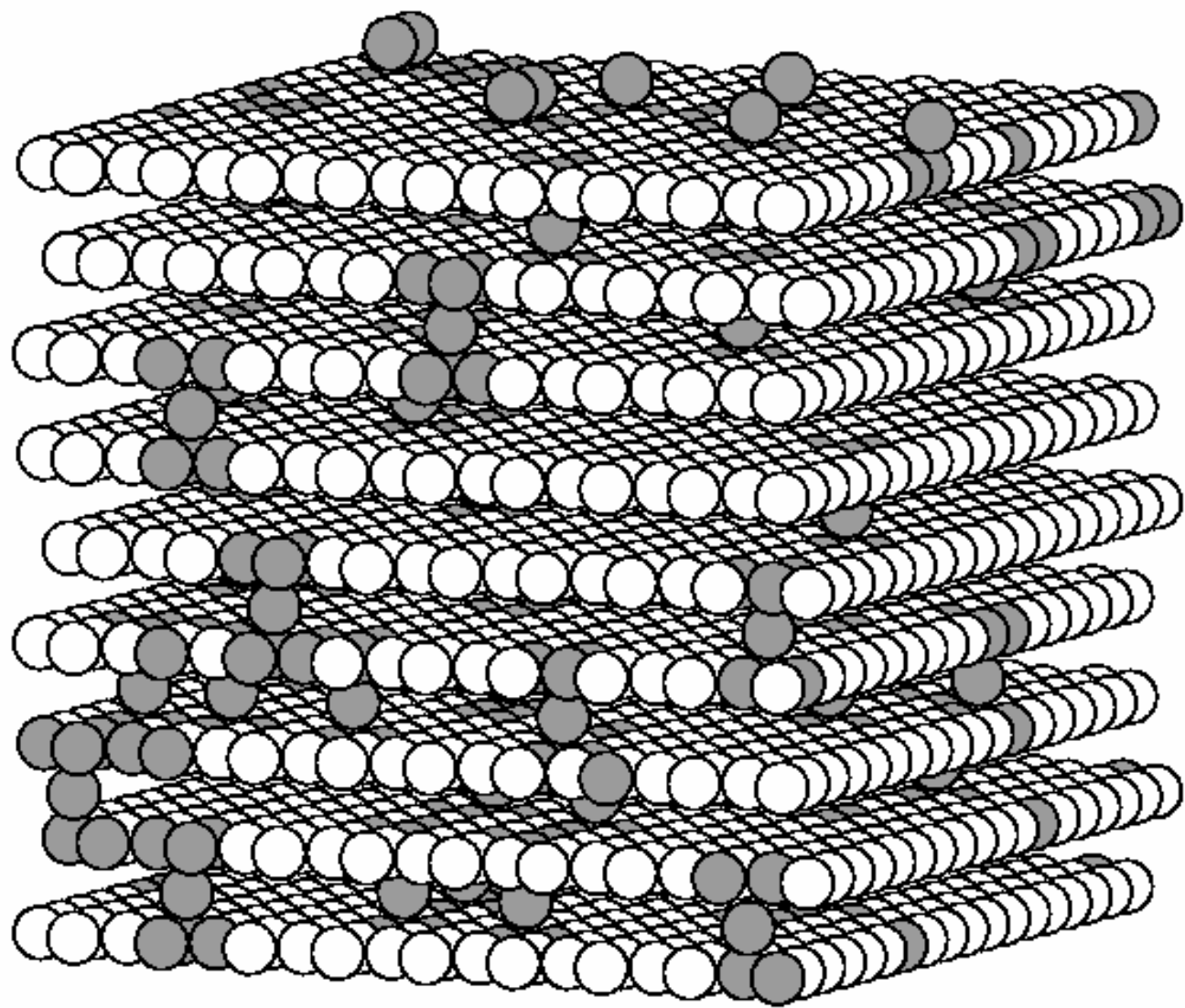
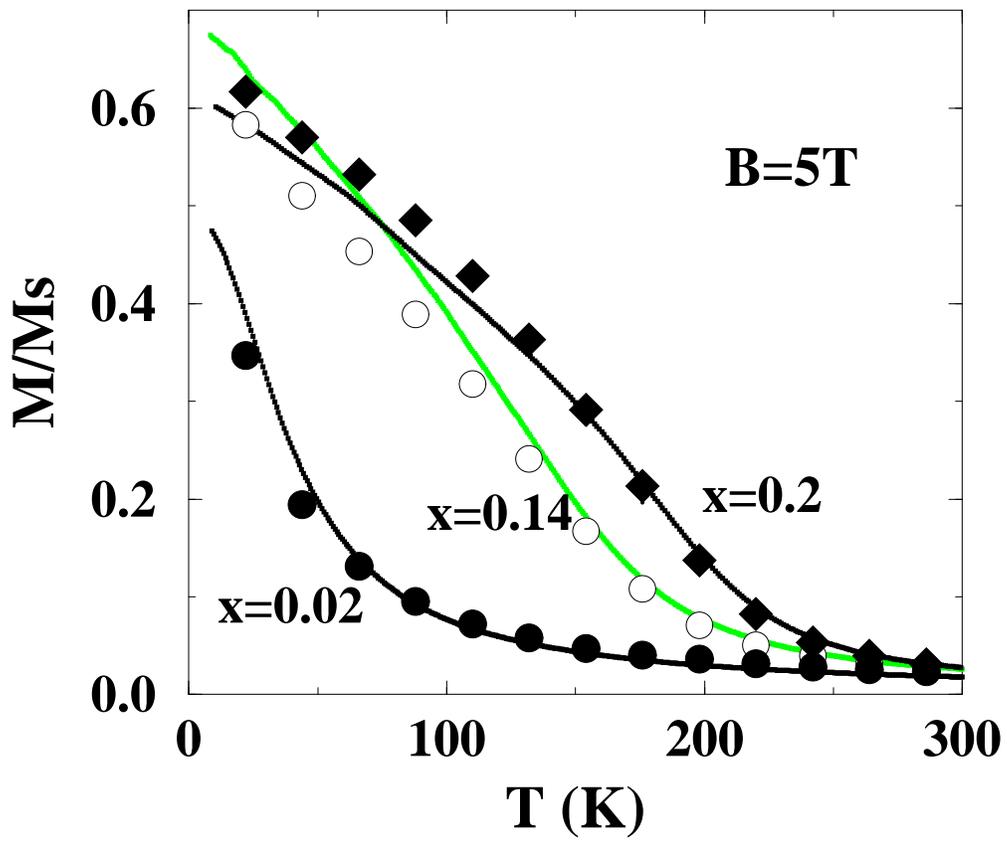
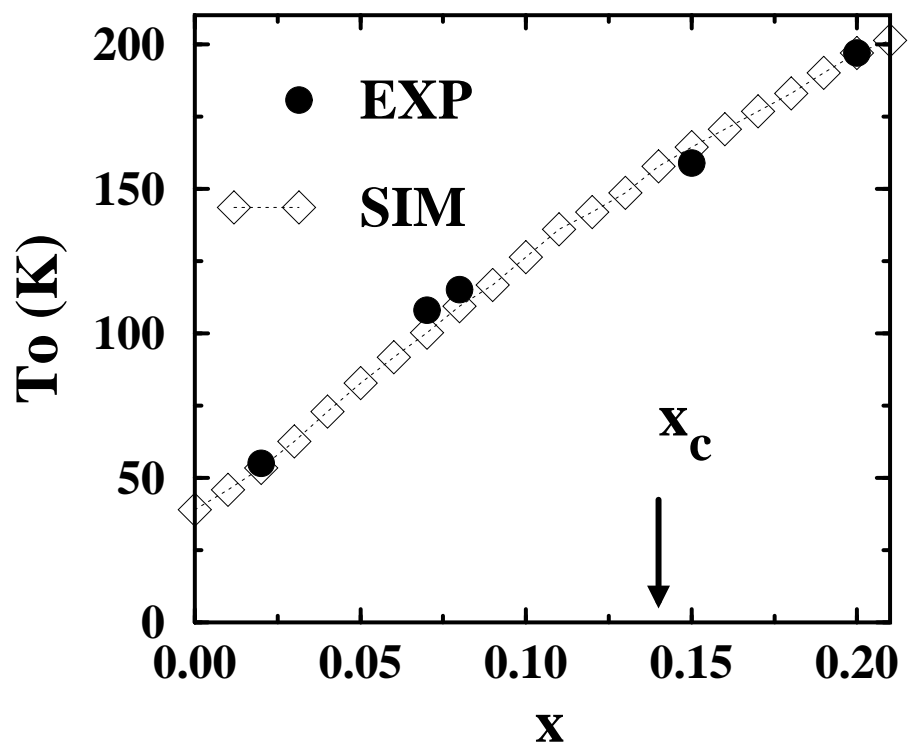
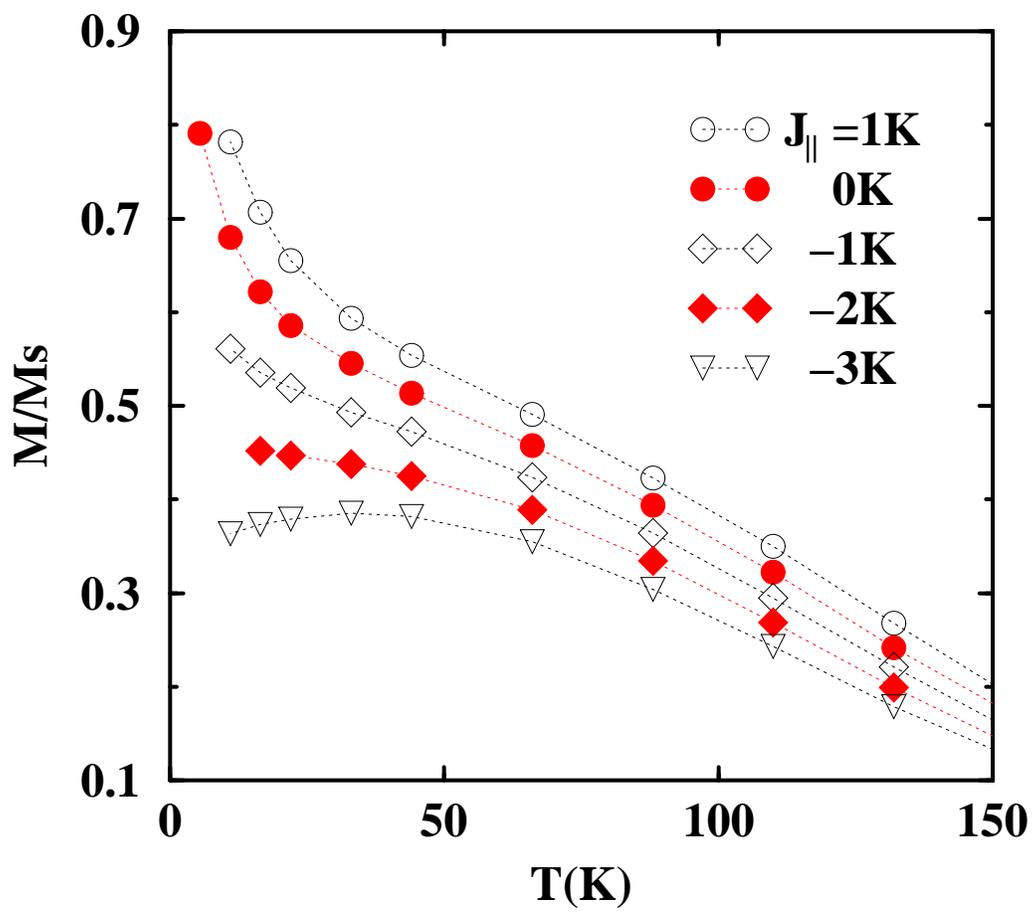
\end{document}